\documentclass[12pt]{article}
\usepackage{amsfonts}
\usepackage{amssymb}
\usepackage{amsmath}
\usepackage{mathrsfs}
\usepackage{bbm}
\usepackage{graphicx}
\usepackage{subfigure}
\usepackage{color}
\usepackage{verbatim}
\usepackage{algorithm}
\usepackage{amsmath}
\usepackage{algpseudocode}
\usepackage{setspace}
\usepackage{geometry}             
\usepackage{epstopdf}
\usepackage{cases}
\usepackage{cite}

\definecolor{rossoCP3}{cmyk}{0,.88,.77,.40}

\usepackage[hyperindex=true,
pdfstartview=FitH,
bookmarksnumbered=true,
bookmarksopen=true,
citecolor=blue,
linkcolor=blue,
colorlinks=true,
urlcolor=rossoCP3,
unicode]{hyperref}

\begin{document}

\title{\bf
	Thermodynamic geometry of the RN-AdS black hole and non-local observables}
{\author{\small Chao Wang${}^{1,2}${}, Bin Wu${}^{1,2,3,4}$\thanks{{\em email}: \href{mailto:binwu@nwu.edu.cn}{binwu@nwu.edu.cn}}{ }, Zhen-Ming Xu${}^{1,2,3,4}${}, and Wen-Li Yang${}^{1,2,3,4}$
		\vspace{5pt}\\
		\small $^{1}${\it School of Physics, Northwest University, Xi'an 710127, China}\\
		\small $^{2}${\it Institute of Modern Physics, Northwest University, Xi'an 710127, China}\\
		\small $^{3}${\it Shaanxi Key Laboratory for Theoretical Physics Frontiers, Xi'an 710127, China}\\
		\small $^{4}${\it Peng Huanwu Center for Fundamental Theory, Xi'an 710127, China}
	}
}

\date{}
\maketitle

\begin{spacing}{1.2}
\begin{abstract}
 In this paper, we show the relation between the thermodynamic geometry of a four-dimensional Reissner-Nordström-AdS (RN-AdS) black hole and non-local observables in boundary field theory. Instead of introducing the critical point associating with the black hole charge to nondimensionalize the thermodynamics parameters, we use the cosmological constant to rescale these variables, so that a universal specific equation of state of the black hole is obtained. Further, the correspondence between thermodynamic properties of the black hole and the oscillating behaviors of the non-local observables has been studied numerically. Our results indicate that the study of the dual field theory will reveal to us the thermodynamic geometry of the AdS black hole.
\end{abstract}

\section{Introduction}

The pioneering work by Hawking and Bekenstein about the black hole temperature and entropy \cite{Bekenstein:1973ur,Hawking:1974sw} made people realize that black hole (BH) should also be a thermodynamic system, which exhibits the phase transition behavior. After the discovery of the correspondence between the Hawking-Page phase transition \cite{Hawking:1982dh} and the confinement/deconfinement phase transition \cite{Witten:1998zw} in the dual field theory, the phase transition becomes more charming. Especially the Van der Waals-like phase transition for it reveal the connection between black holes and ordinary thermodynamic systems \cite{Chamblin:1999tk}. More precisely, as the charge increases, the black hole will undergo the second-order and first-order phase transition successively before it reaches the stable phase. 

The Van der Waals phase transition was reconstructed in the extended phase space \cite{Kubiznak:2012wp,Kastor:2009wy,Dolan:2011xt,Cvetic:2010jb}, with the negative cosmological constant $\Lambda=-(d-1) (d-2)/l^2$ being treated as the thermodynamic pressure $P=-\Lambda/8\pi G$ and its conjugating quantity is the thermodynamic volume $V$, here $l$ is the AdS radius. Within this framework, the phase transition and critical behavior of black holes have been numerously studied in \cite{Kubiznak:2016qmn, Toledo:2019amt, Hendi:2012um, Wei:2012ui, Cai:2013qga, Zhao:2013oza, Altamirano:2013ane, Spallucci:2013osa, Xu:2014tja}. However, due to the importance of the AdS black hole in the Anti-de Sitter/Conformal Field Theory (AdS/CFT) duality, which describe the correspondence between the quantum gravity and the gauge field theory resides in the boundary \cite{Maldacena:1997re,Gubser:1998bc}, the holographic interpretation of the variation of $\Lambda$ attracted much attention. 

Several works in \cite{Dolan:2014cja,Kastor:2014dra,Karch:2015rpa} have proposed that the variation of $\Lambda$ corresponds to varying the number of color $\mathcal{N}$, alternatively the number of degrees of freedom $ \mathcal{N}^2$. In CFT, the number of degrees of freedom is denoted by central charge with $C= l^{d-2}/G_d$ \cite{Henningson:1998gx,Freedman:1999gp,Myers:2010xs}, where $G_d$ is the Newtonian constant in $d$-dimensional spacetime. Therefore, the change in $\Lambda$ corresponds to varying central charge. Or if requiring a fixed CFT, it is suggested that the Newtonian constant $G_d$ should vary as the varying of $\Lambda$ to keep the central charge $C$ as a constant \cite{Cong:2021fnf}. Recently, Visser derived the holographic thermodynamics in the dual field theory by applying the central charge as a thermal variable, which plays a similar role as the amount of matter in the thermodynamic system\cite{Visser:2021eqk}. Inspired by this, much work has been done to reveal the properties of the black hole in the context of varying Newtonian constant\cite{Rafiee:2021hyj,Cong:2021fnf,Cong:2021jgb}. A step further, in research \cite{Zeyuan:2021uol,Gao:2021xtt,Wang:2021cmz,Zhao:2022dgc}, by fixing the AdS radius, a novel thermodynamic pair $C$ and $\mu$ related to varying $G_d$ are introduced, consequently the additivity of the black hole thermodynamics has been developed.

While the development of black hole thermodynamics is in full swing, how to probe their thermodynamic behavior has attracted much interest. Driven by this, the detection of the thermodynamic behavior of the RN-AdS black hole by non-local observables in dual field theory has been investigated \cite{Johnson:2013dka}, and the result demonstrated that there exists an oscillating behavior in the temperature-holographic entanglement entropy \cite{Ryu:2006bv,Ryu:2006ef} plane, which resembles the Van der Waals phase transition. Furthermore, they examined the critical behavior and the Maxwell equal law \cite{Caceres:2015vsa,Nguyen:2015wfa}, and they are all fulfilled. The further study implied that the oscillating behavior was also found in the coordinate space organized by the Hawking temperature and geodesic length on the AdS boundary which is related to the equal time two-point correlation function and Wilson loop \cite{Zeng:2015wtt,Zeng:2016fsb}.

The phase transition in classical thermodynamics originates from intermolecular interaction, but the microstructure of black holes is still a mystery. The introduction of the Ruppeiner geometry \cite{PhysRevA.24.488,Ruppeiner:1983zz,Ruppeiner:1995zz} provides some insight into that. Considering the fluctuation theory, line elements are proposed to measure the distance between fluctuation states, which has the form of
\begin{equation}
	\Delta l^2= - \frac{\partial^2 S}{\partial X^\mu \partial X^\nu} \Delta X^\mu X^\nu = g^R_{\mu \nu} \Delta X^\mu \Delta X^\nu.    \label{Rlineelement}
\end{equation}
Here $S$ is the entropy of the system, and $X^\mu$ is the thermodynamic coordinates depending on the choice of the thermodynamic differential relation. The curvature scalar calculated from Eq.(\ref{Rlineelement}) denotes the interaction between the adjacent part of the fluid system.  Especially speaking, the curvature scalar with the positive (negative) value implied the interaction inside the black hole is a repulsive (attractive) domain, and the noninteracting system corresponding to the flat Ruppeiner metric \cite{Ruppeiner:2008kd,Ruppeiner:2010,Sahay:2010wi,Wei:2015iwa,Zhang:2015ova,Wei:2017icx,Chaturvedi:2017vgq,Wei:2019uqg,Wei:2019yvs,Xu:2020ftx,Wang:2021vbn,Wang:2021llu}. In this sense, the application of the Ruppeiner geometry is the reverse process of statistical physics, that is, detecting the microstructure of a system with its thermodynamic behavior. It is also found that the divergent point of the curvature scalar corresponds to the phase transition point \cite{Ruppeiner:2011gm}, which implied the phase structure will be exposed by the curvature scalar.

Considering that the non-local variables have the same oscillating behavior as the black hole phase transition, it is natural to ask whether the information of the underlying microstructure depicted by the thermodynamic geometry can be read in a given CFT. Since we are considering a fixed CFT, so the central charge of the CFT is fixed, as well as the $l$ and $G_d$ are kept as constants, which corresponds to a particular thermodynamic state of the black hole.  Inspired by this, we adopted the connection between the black hole entropy and the non-local observables in dual field theory, including the holographic entanglement entropy and the two-point correlation function, the result specified that the observation of the quantified in CFT will expose the information of the black hole phase structure and thermodynamic geometry.

The outline of this paper is as follows. In Sec.\ref{II}, the phase structure and the Ruppeiner of the RN-AdS black hole have been reviewed. The numerical result of the holographic entanglement entropy and two-point correlation function of the RN-AdS black hole will be investigated in Sec.\ref{III}. We end this paper with a conclusion in Sec.\ref{IIIII}. Throughout this paper, we adopt the units $\hbar=c=k_B=G=1$ for convenience.

\section{RN-AdS Black Hole}\label{II}

The RN-AdS black hole in $4$-dimensional spacetime is characterized by the action in the form of 
\begin{equation}
	I=\frac{1}{16 \pi} \int \sqrt{-g} \mathrm{d} x^{4}\left(R-F^{\mu \nu} F_{\mu \nu}+\frac{6}{l^{2}}\right),
\end{equation}
where $l$ is the AdS radius, and the equation of motion is
\begin{align}
	\mathrm{d}s^2 &= -f(r) \mathrm{d} t^2 + \frac{1}{f(r)} \mathrm{d} r^2 + r^2 \left(\mathrm{d}\theta^2+\sin^2 \theta \mathrm{d}\phi^2 \right),    \nonumber\\
	&F_{\mu \nu}=\partial_{\mu} A_{\nu}-\partial_{\nu} A_{\mu}, \quad A_\mu= \left(  -\frac{Q}{r},0,0,0  \right).    \nonumber
\end{align}
The metric function $f(r)$ can be obtained easily
\begin{equation}
	f(r)=1-\frac{2 M}{r}+\frac{Q^{2}}{r^{2}}+\frac{r^{2}}{l^{2}}.   \label{f}
\end{equation}
The parameters $M$ and $Q$ in the metric function are the mass and charge of the black hole, respectively. The temperature and entropy are given by
\begin{align}
	&T =\frac{1}{4 \pi r_h}\left(1+\frac{3 r_h^{2}}{l^{2}}-\frac{Q^{2}}{r_h^{2}}\right), \label{temperature}\\
	&S=\pi r^2_h,  \label{entropy}
\end{align}
where $r_h$ is the position of the outer event horizon. The mass of the black hole can be deduced with the condition $f(r_h)=0$
\begin{equation}
	M=\frac{r_h}{2}+\frac{\Phi Q}{2}+\frac{r_h^{3}}{2 l^{2}},
\end{equation}
here $\Phi=Q/r_h$ is the electric potential difference between the horizon and infinity. 

\begin{figure}[!h]
	\centering
	\includegraphics[width=7cm]{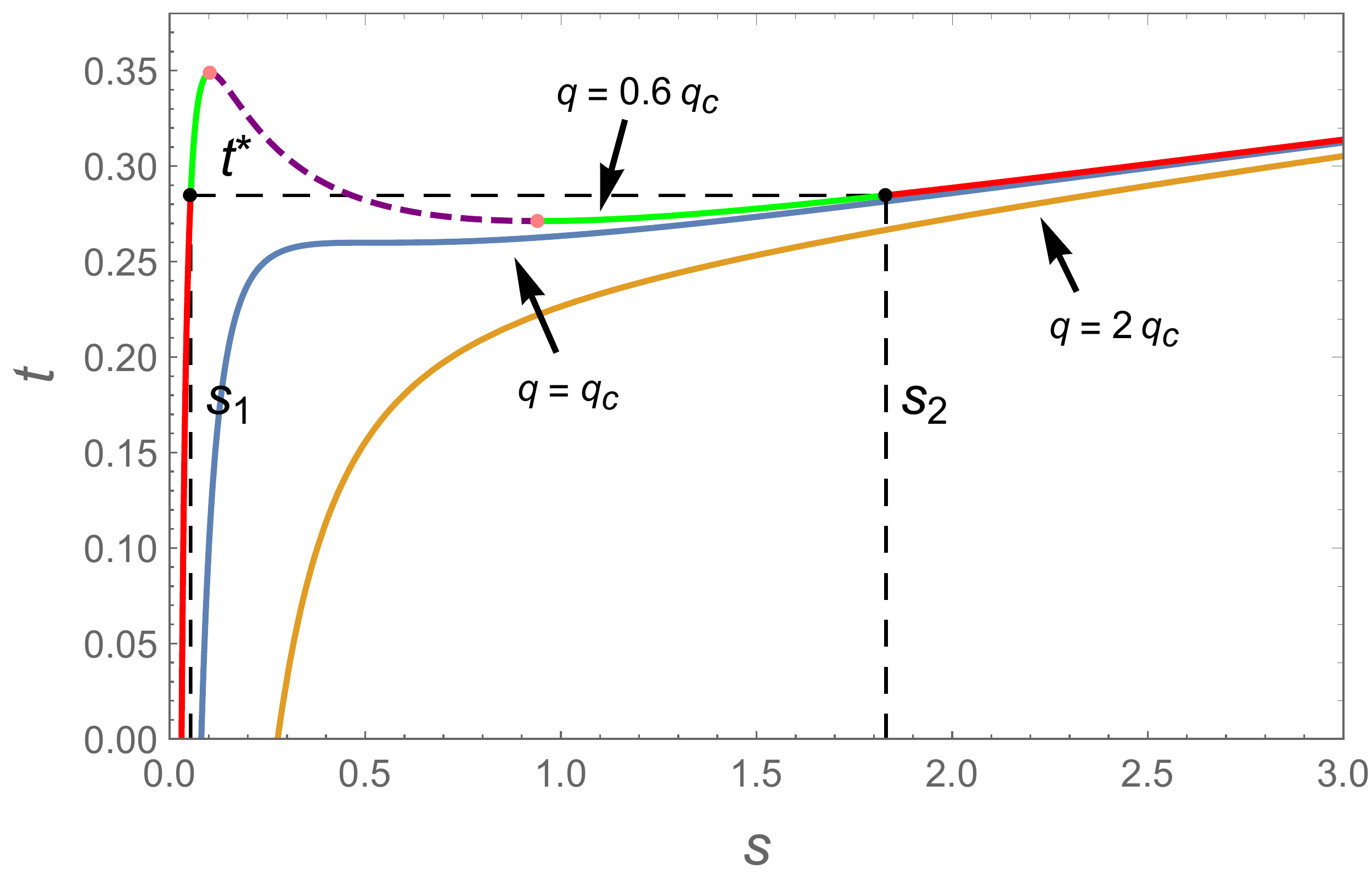}
	\caption{Iso-$q$ curves in $\{t,s\}$ coordinate space. Here we set $q=0.6 q_c$, $q_c$ and $2 q_c$ from top to bottom. The maxwell equal area law is constructed by the black dashed line.}\label{T-S}
\end{figure}

Next, we would like to analysis the small/large phase transition of the RN-AdS black hole. Usually, one would introduce the reduced parameters by the ratio of the variables to the corresponding critical value, which satisfied $\left( \frac{\partial T}{\partial S } \right)_Q = \left( \frac{\partial^2 T}{\partial S^2 }  \right)_Q =0$. According the equation of state $P=\frac{T}{v}-\frac{1}{2 \pi v^{2}}+\frac{2 Q^{2}}{\pi v^{4}}$ in \cite{Kubiznak:2012wp} with $v=2r_h$, the critical point is denoted by
\begin{equation}
	P_{\mathrm{c}}=\frac{1}{96 \pi Q^{2}}, \quad T_{\mathrm{c}}=\frac{\sqrt{6}}{18 \pi Q},  \quad v_{\mathrm{c}}=2 \sqrt{6} Q, \nonumber
\end{equation}
and the specific equation of state in reduced parameters can be written as
\begin{equation}
	\tilde{P}=\frac{8 \tilde{T}}{3 \tilde{v}}-\frac{2}{\tilde{v}^{2}}+\frac{1}{3 \tilde{v}^{4}}, \label{EoS}
\end{equation}
where $\tilde{P}= P/P_c$, $\tilde{T}=T/T_c$ and $\tilde{v}=v/v_c$. We see that the charge $Q$ disappear in the equation of state Eq.(\ref{EoS}), which means the information about the black hole charge is missing during this operation. Since for van der Waals fluids, the critical point is labeled by its model parameters, and the introduction of the reduced parameter leads to the law of corresponding states. We notice that the cosmological constant play the role of model parameter as that of van der Waals fluids, according to the dimensions of the thermodynamic variable, the reduced dimensionless  thermal parameters are introduced as
\begin{equation}
	s=S/l^2, \quad q=Q/l, \quad t=Tl.  \label{rp}
\end{equation}
In this way, the specific state equation with the dimensionless parameters will hold for all charged AdS black holes, who have the form of
\begin{equation}
	t= \frac{3 s^2 + \pi s - \pi^2 q^2}{4 (\pi s)^{3/2}}. \label{rT}
\end{equation}
The rewritten equation of state does not explicitly contain the AdS radius $l$, thus $l$ can be set as an arbitrary constant. The critical point decided by $\left( \frac{\partial t}{\partial s } \right)_q = \left( \frac{\partial^2 t}{\partial s^2 }  \right)_q =0$ is marked as
\begin{equation}
	s_c=\frac{\pi}{6}, \quad q_c=\frac{1}{6}, \quad t_c=\frac{\sqrt{6}}{3 \pi}. \nonumber
\end{equation}
Based on the Eq.(\ref{temperature}), Eq.(\ref{entropy}) and Eq.(\ref{rp}), we show the iso-$q$ process in Fig.\ref{T-S} with $q=0.6 q_c$, $q_c$ and $2 q_c$ from top to bottom, 
The box formed by the black dashed line corresponds to the Maxwell equal area law of the Van der Waals-like phase transition. The coexistence temperature $t^*$ and $s_{1,2}$ are the coordinates of the box vertices, which can be obtained as
\begin{align}
	&t^*=\frac{1}{\pi}\sqrt{1-2q},    \label{Tstar}\\
	&s_{1,2}=\frac{\pi}{4} (\sqrt{1-2 q} \pm \sqrt{1-6q})^2.  \label{S12}
\end{align}
While the charge is less than $q_c$, the iso-$q$ curve exhibits oscillatory behavior. The thermodynamics theory tells us that a small stable hole will directly transform into a large stable hole when the temperature of the black hole exceeds $t^*$. Together with the local extremal point that paint in purple, the iso-$q$ curve was divided into five-segment. The red solid lines correspond to the small (SBH) and large black holes (LBH), which are stable. The black hole on the purple dashed line is thermally unstable, and at this point the black hole can be in either a large or small black hole state (SBH$+$ LBH). The green solid lines are the metastable curves of the black hole, and they separate the stable and unstable states of the black hole, with the left and right segments corresponding to superheated small BH (SHSBH) and supercooled large BH (SCLBH), respectively. As the charge reaches the critical value, these two local extremal points merge into one at the inflection point in the orange curve in Fig.\ref{T-S}. Further increasing the value of $q$, the black hole is in the supercritical phase (SCBH).

\begin{figure}[!h]
	\centering
	\subfigure[]{
		\includegraphics[width=4.75cm]{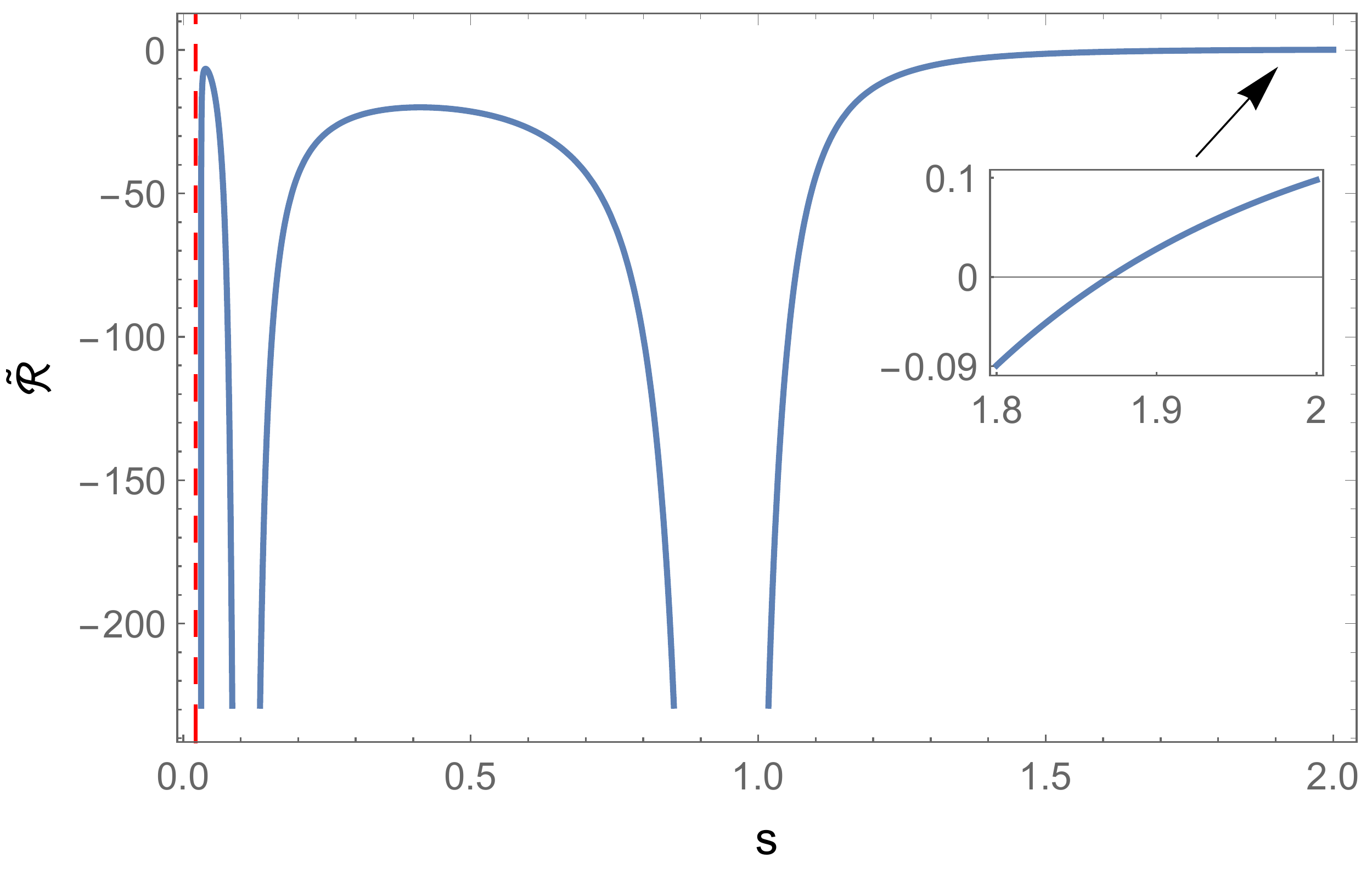}
	}
	\quad
	\subfigure[]{
		\includegraphics[width=4.75cm]{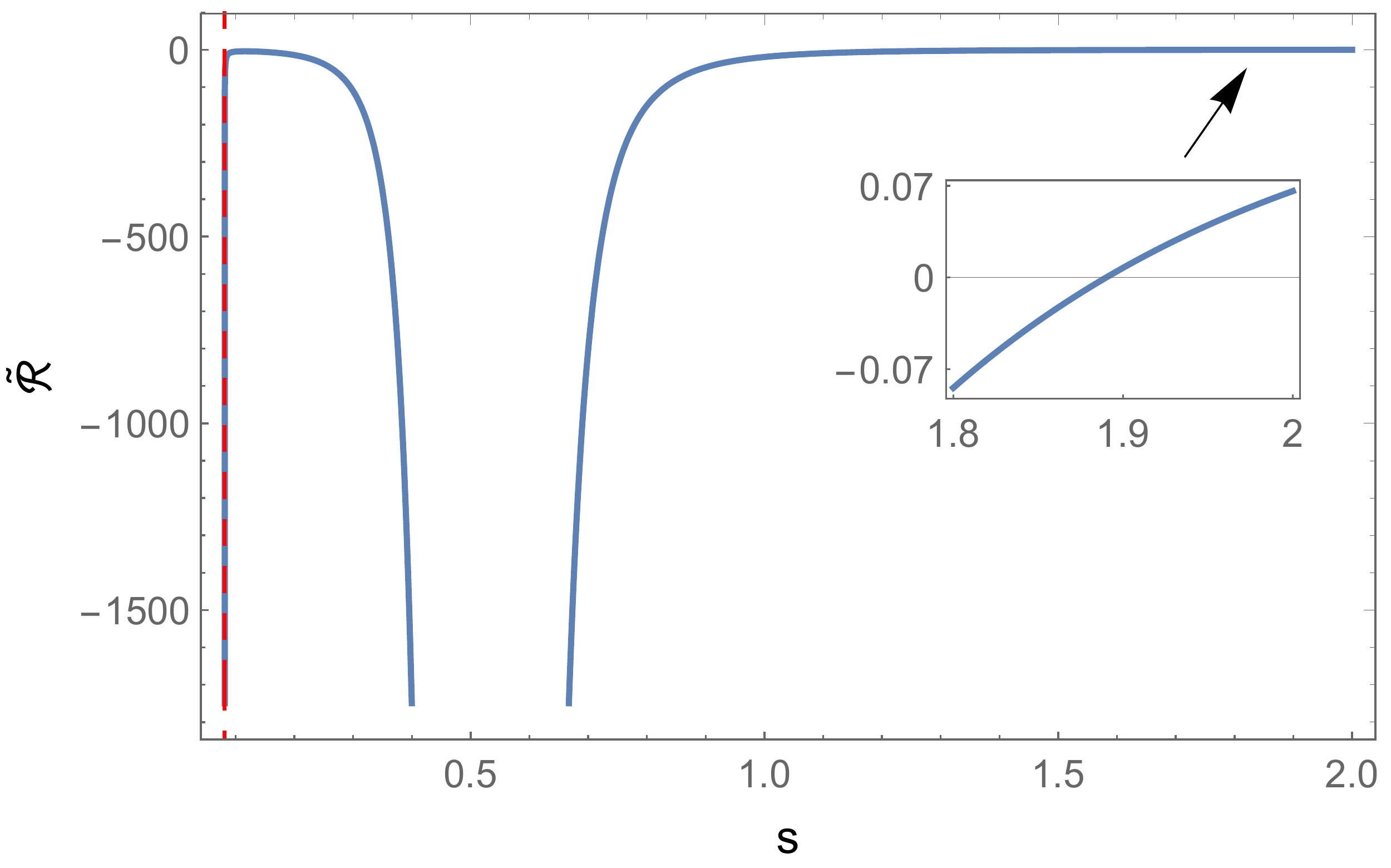}
	}
	\subfigure[]{
		\includegraphics[width=4.6cm]{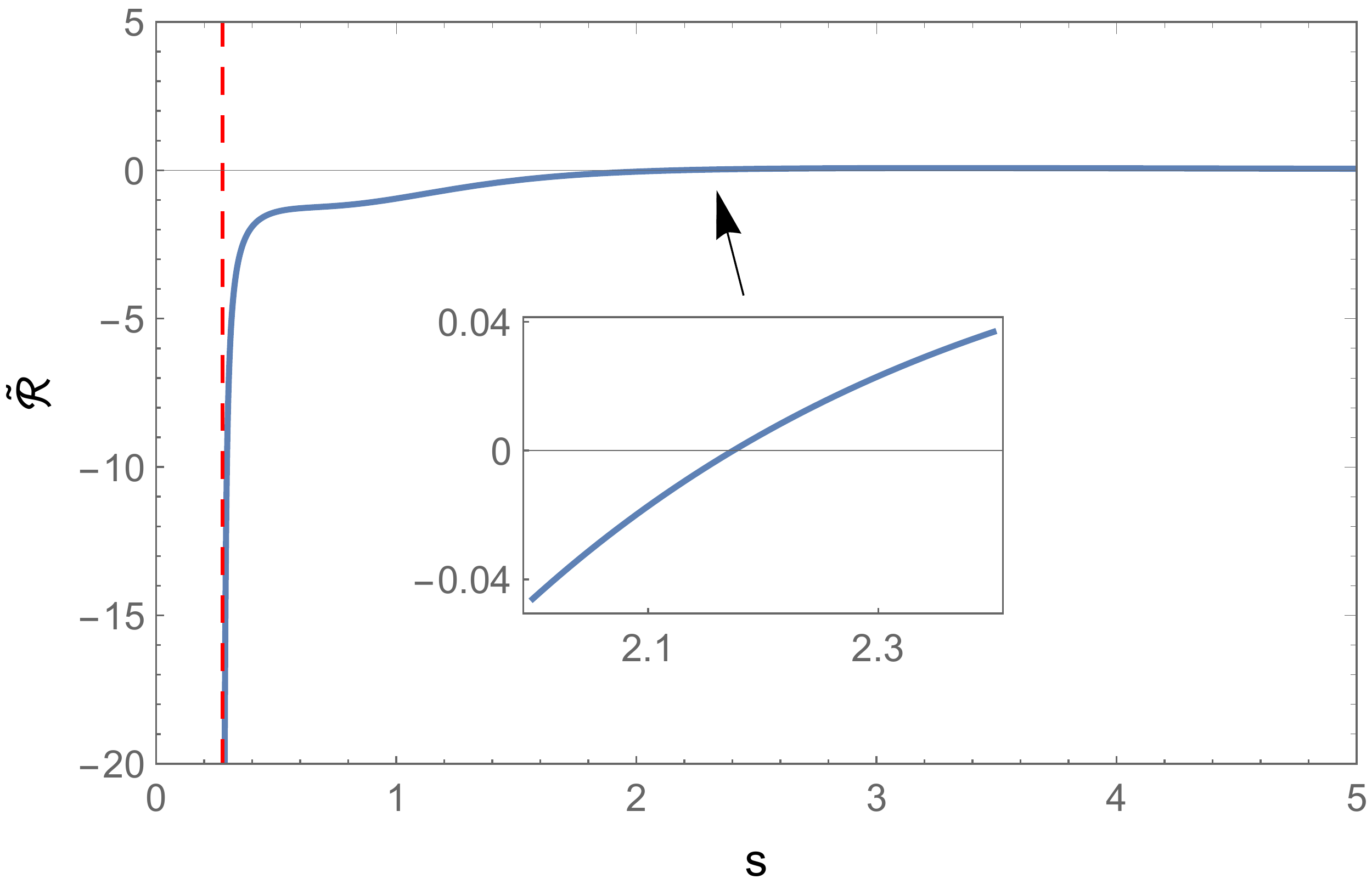}
	}
	\caption{The dimensionless curvature scalar for RN-AdS black hole in terms of $s$. Here we set $q=0.6 q_c$, $q_c$ and $2q_c$ from left to right. The entropy of the red dotted line corresponds to the vanishing value of Hawking temperature Eq.(\ref{temperature}).}\label{R-SQ}
\end{figure}

To reveal the information on the underlying microstructure of the black hole, we adopted the Ruppeiner geometry of the RN-AdS black hole with its internal energy $U=M-\Phi Q$\cite{Zhang:2015ova}, which is the representation of the intrinsic properties of the Charged AdS-black hole. The thermodynamic curvature scalar comes to
\begin{align}
	\mathcal{R}= -\frac{(Q^2-r^2_h)^2 + 3 r^2_h (10 Q^4 -9 Q^2 r^2_h + 3 r^4_h)/l^2 + 18 r^6_h (3 Q^2 - r^2_h)/l^4}
	         {\pi (r^2_h - Q^2 +3 r^4_h /l^2) (3 Q^2 - r^2_h +3 r^4_h / l^2)^2},  \label{R}
\end{align}
with dimensiom $[\mathcal{R}]=1/[l]^2$. Combine the equations in Eq.(\ref{temperature}) and Eq.(\ref{entropy}), we show the vary of dimensionless curvature scalar $\widetilde{\mathcal{R}}=\mathcal{R}l^2$ in terms of entropy $s$ with $q=0.6 q_c$, $q_c$ and $2q_c$ in Fig.\ref{R-SQ}, respectively. 
From Eq.(\ref{temperature}), we know that the scalar curvature starts from a negative infinity caused by $t=0$ marked by the red dotted line, which describes the extremal black holes. When $q<q_c$, the scalar curvature is divided into three-part by two divergent points, they are the stable small, unstable and stable large phases from left to right. With the increase of charge, these divergent points get close and merge into one at $q=q_c$, and the black hole is undergone a second-order phase transition. With the increasing of charge, the curve will be continuous and the black hole is in the supercritical phase. The figure also implied that the scalar curvature is always negative (implying the interaction in the black hole is an attractive domain) until entropy is large enough.

\begin{figure}[!h]
	\centering
	\includegraphics[width=7cm]{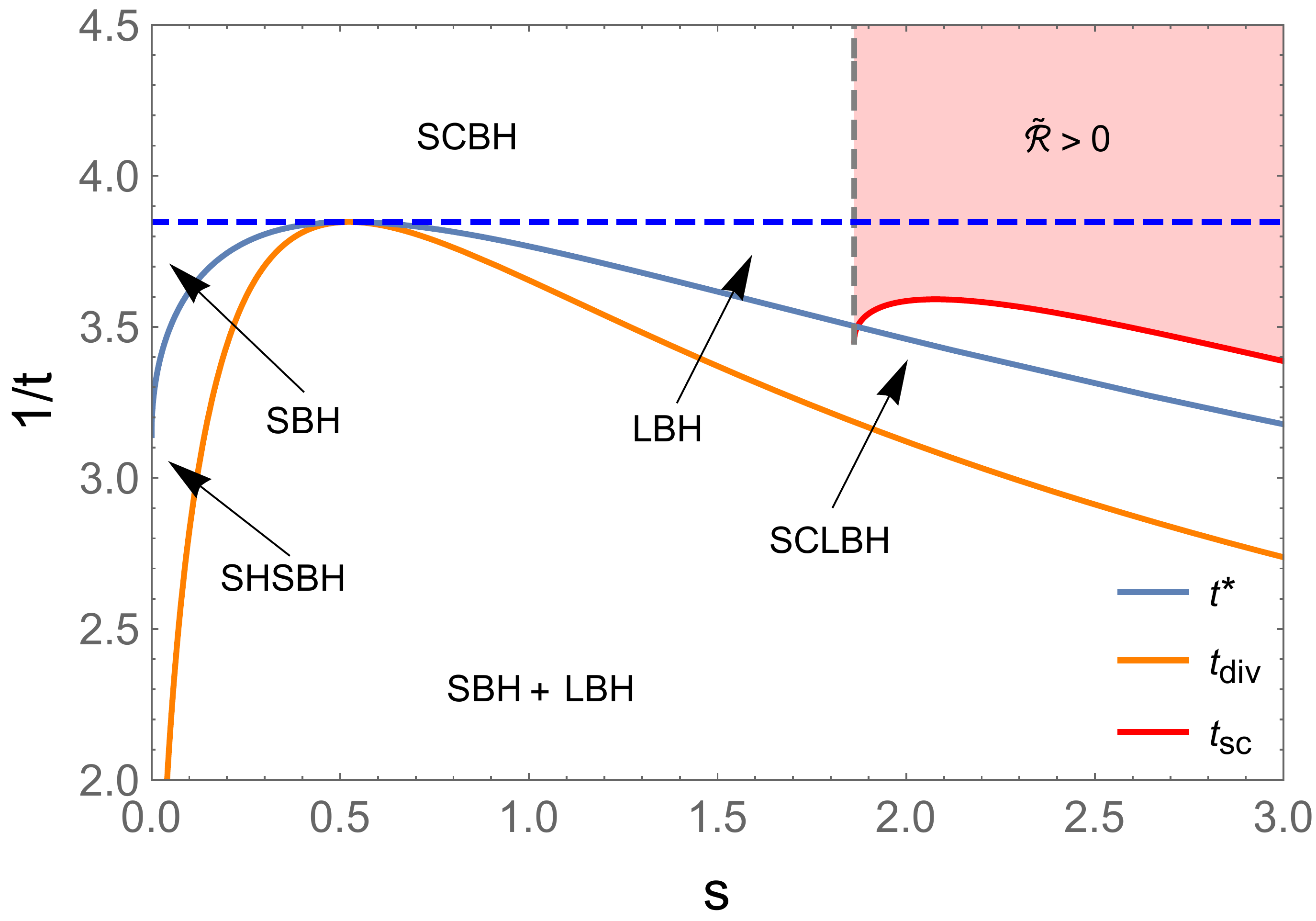}	
	\caption{The characteristic curve of scalar curvature in coordinate space $\{t,s\}$. The orange and red are solid lines are the divergence curve and the variable sign curve of the curvature scalar, respectively. The blue solid line is the coexistence curve in Eq.(\ref{Tstar}). The blue dashed curve in the figure corresponds to the critical point temperature.}\label{R-TS}
\end{figure}

As discussed above, the vanishing point and divergent point of $\widetilde{\mathcal{R}}$ are helpful for us to learn the underlying microstructure and the phase structure of the black hole, checking the characteristic curve including the sign-changing curve $t_{\mathrm{sc}}$ and the divergent curve $t_{\mathrm{div}}$ of scalar curvature is important. With Eq.(\ref{temperature}), Eq.(\ref{entropy}) and Eq.(\ref{R}), we plot these characteristic curves in Fig.\ref{R-TS}.
The blue solid curve in the figure is the coexistence curve based on Eq.(\ref{Tstar}) and Eq.(\ref{S12}), which is divided into the saturated small phase and saturated large phase by the extreme point. The orange and red curves are the divergence curve and the sign-changing curve of $\widetilde{\mathcal{R}}$, respectively. The area above $t_{\mathrm{sc}}$ curve that is painted in pink corresponds to the positive value of the curvature scalar, and the interaction is a repulsive domain. While other areas with $\widetilde{\mathcal{R}}<0$ implied the interaction between the parts of the black hole is an attractive domain. The figure tells us that only the large black hole would show the repulsive interaction. The phase structure was also exhibited in the figure. Above the blue dashed line, the black hole was in the supercritical phase.

\section{Thermodynamic geometry and non-local variables}\label{III}

As we have reviewed the thermodynamic geometry of the RN-AdS black hole, we will further investigate its relation to the non-local observables in a given CFT, including the holographic Entanglement Entropy (HEE) and two-point correlation function. We would like to start this topic with HEE, which has been proved to show a similar phase transition and critical behavior as that of the black hole entropy. Therefore, in order to explore how similar they are, we explored whether HEE can reflect the thermodynamic geometry of a black hole. 

The Entanglement Entropy (EE) denotes the relationship between two subsystems of a quantum system, which is denoted by $A$ and $A^c$. When the quantum system lived on a Conformal Field Theory, the entanglement entropy can be computed by the Ryu-Takayanagi (RT) recipe \cite{Ryu:2006bv,Ryu:2006ef}
\begin{equation}
	S_{A}=\frac{\operatorname{Area}\left(\Gamma_{A}\right)}{4 G}.  \label{area}
\end{equation}
Here the $\Gamma_A$ is a codimension-2 minimal surface in the corresponding AdS space, which owns the same boundary condition as $A$. According to the metric function Eq.(\ref{f}), Eq.(\ref{area}) can be written as
\begin{equation}
	S=\frac{\pi}{2} \int_{0}^{\theta_{0}} \mathcal{L}(r(\theta),\theta) d \theta, 
	\quad \mathcal{L}=r \sin \theta \sqrt{\frac{\left(r^{\prime}\right)^{2}}{f(r)}+r^{2}},   \label{EEaction}
\end{equation}
where $r'= \mathrm{d}r/\mathrm{d}\theta$ with $\theta_0$ as the boundary condition of HEE in $\theta$ direction. $\mathcal{L}$ is now introduced as the Lagrangian with $\theta$. The only analytical solution of $r(\theta)$ is pure AdS spacetime in the bulk. In the study of HEE, we will solve the equation of motion in ordinary spacetime numerically with the conditions that
\begin{equation}
	r^ \prime (0)=0, \quad r(0)=r_0. \label{boundary}
\end{equation} 

\begin{figure}[!h]
	\centering
	\subfigure[]{
		\includegraphics[width=4.7cm]{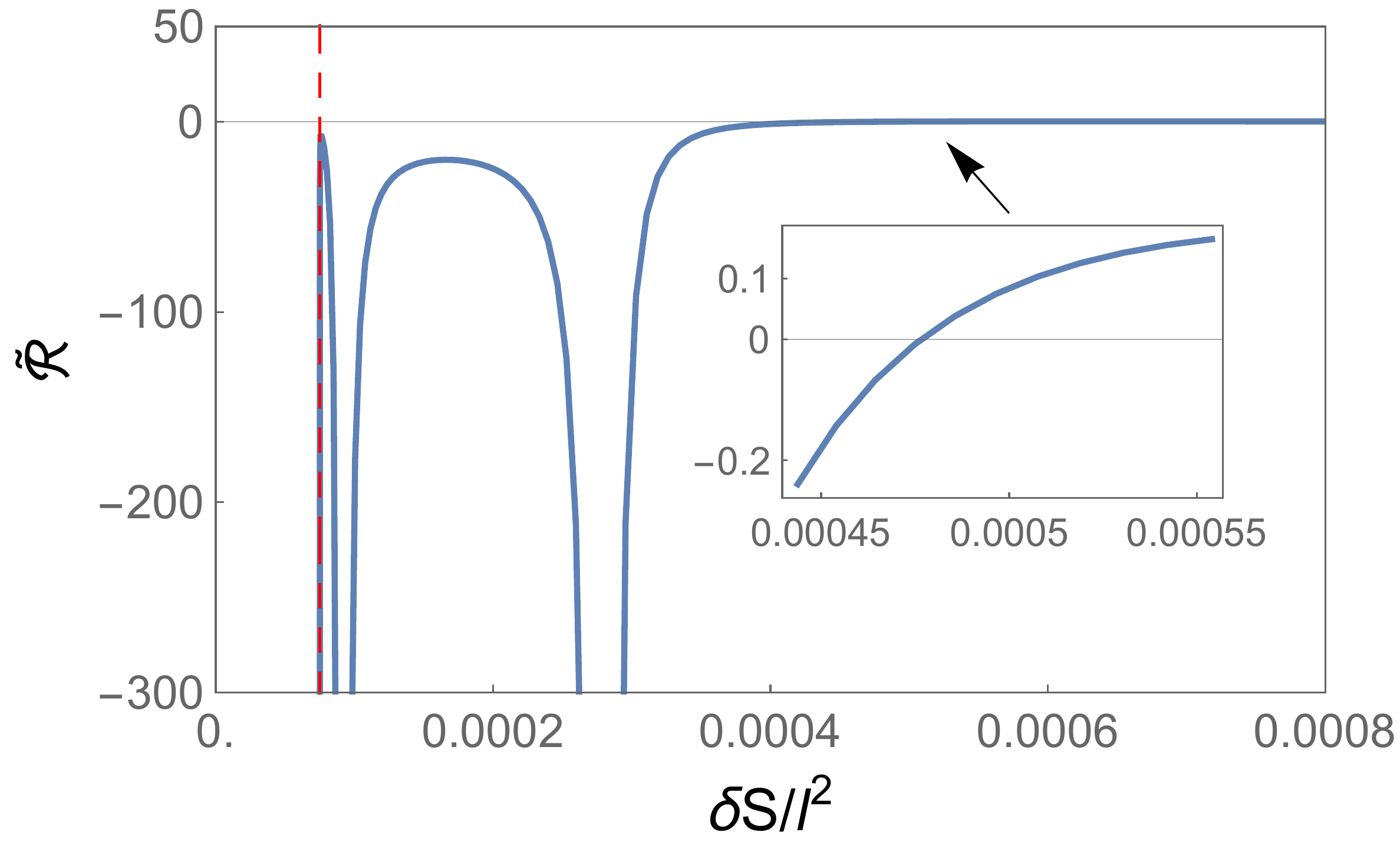}
	}
	\quad
	\subfigure[]{
		\includegraphics[width=4.7cm]{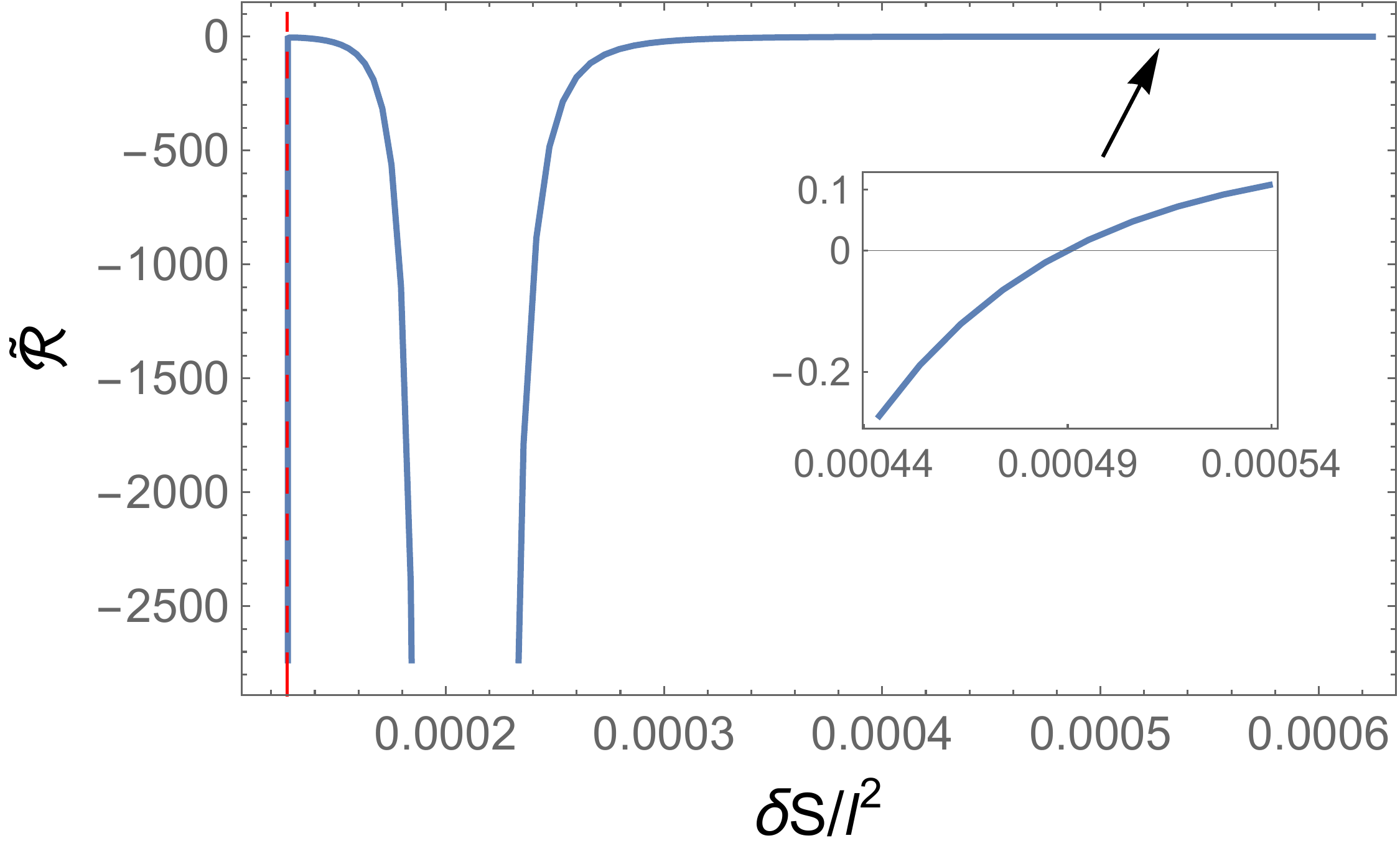}
	}
	\subfigure[]{
		\includegraphics[width=4.7cm]{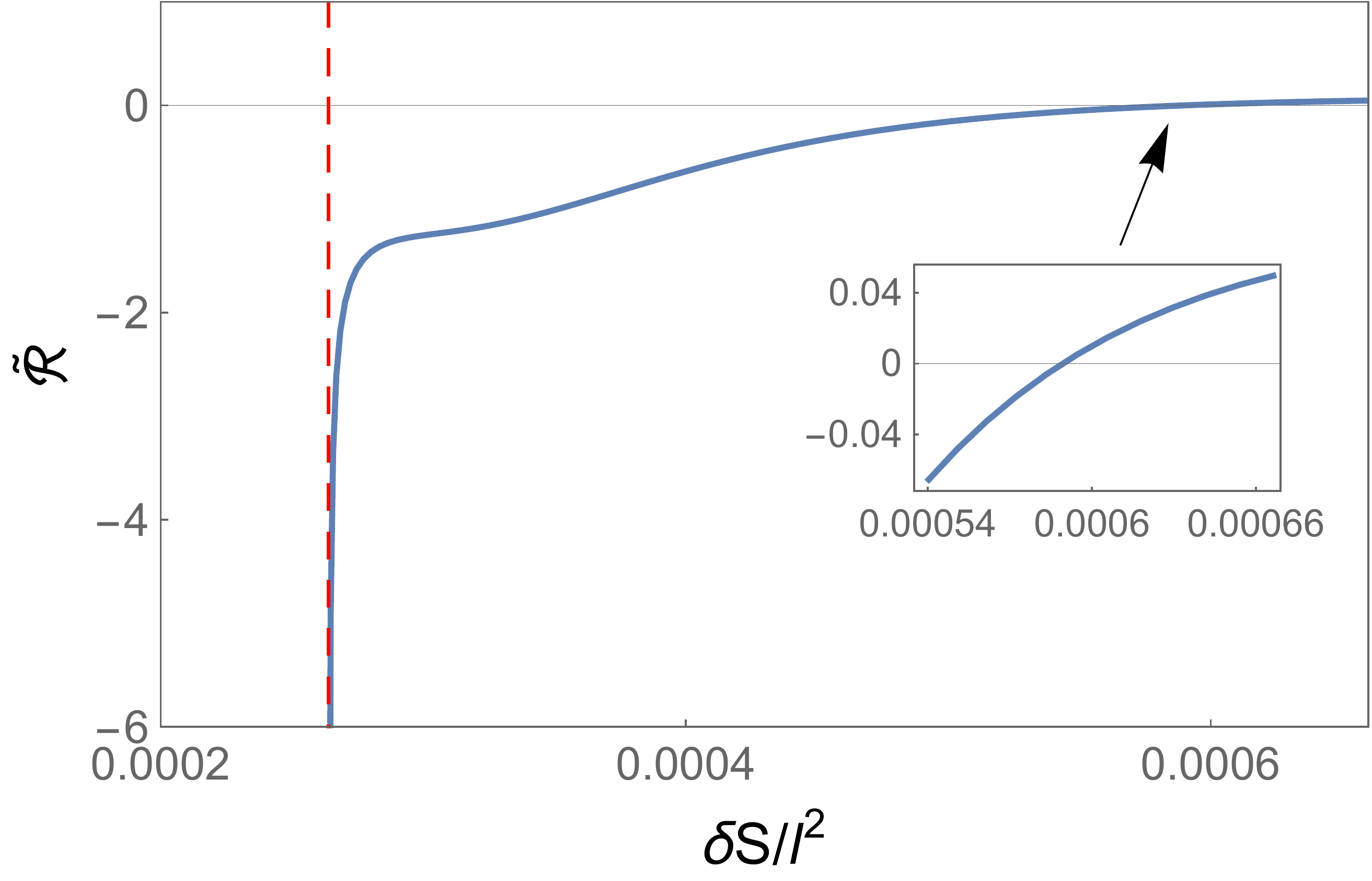}
	}
	\caption{The dimensionless curvature scalar for RN-AdS black hole in terms of $\delta s/l^2$. Here we set  $q=0.6 q_c$, $q_c$ and $2q_c$ from left to right. The entropy at the red line corresponds to the vanishing value of the reduced temperature $t$.}\label{R-HEE}
\end{figure}

Notice that for the UV-divergent of the entanglement entropy, we should regulate it by subtracting the area of the minimal surface in pure AdS, which we denoted as $\delta S$. When calculating the entanglement entropy with RT formulate, we ask $\theta_0$ to be a small value to make sure the minimal surface can return to the subsystem continuously. Therefore we set $\theta_0=0.1$ and UV-cutoff in CFT with $r(0.099)$. To check whether the holographic entanglement entropy can be exploited to reflect the curvature scalar of the black hole, we study the relationship between the Ruppeiner geometry and HEE. The equation of HEE in Eq.(\ref{EEaction}) indicated the relation in $r_h$ and $S_A$, with which we can establish a one-to-one point correspondence between the black hole phase transition and the oscillating behavior of non-local observables in the given CFT. The numerically results of curvature scalar in terms of HEE with different charges can be shown in Fig.\ref{R-HEE}. 
The behavior of the curvature scalar with respect to dimensionless parameter $\delta S/l^2$ with $q=0.6q_c$, $q_c$ and $2q_c$ from left to right. With the numerical method, there is a point-to-point correspondence to that in Fig.\ref{R-SQ}. As the same as the black hole entropy, the entanglement entropy also reveals the phase structure of the RN-AdS black hole with a fixed charge, which is identical to that of the Van der Waals fluid. What's more, with a glimpse of the value of HEE, we can assert the interaction within the corresponding black hole. 

\begin{figure}[!h]
	\centering
	\includegraphics[width=7cm]{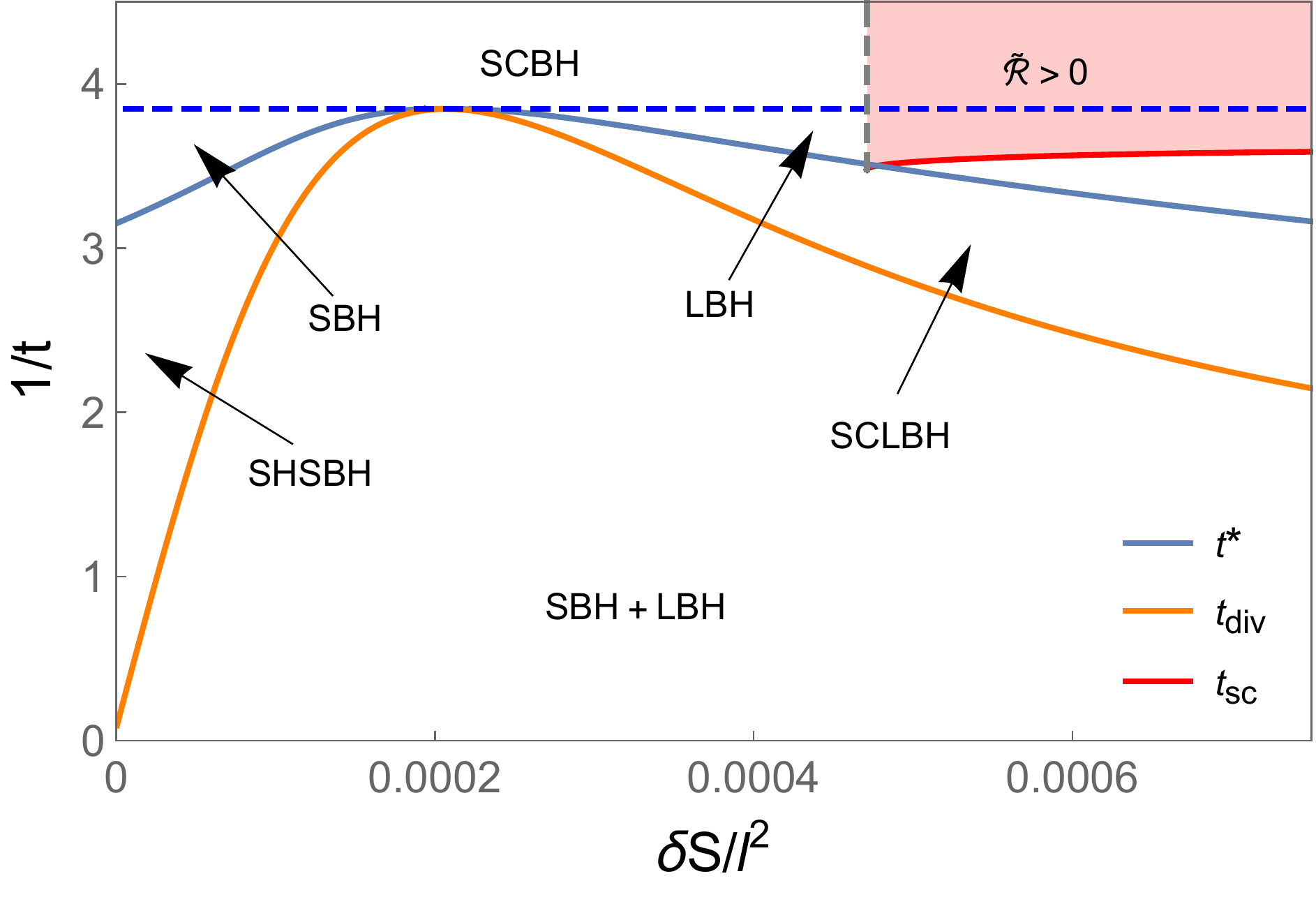}
	\caption{The characteristic curve of curvature scalar in coordinate space $\{\delta S/l^2, 1/t\}$. The orange and red are solid lines are the divergence curve and sign-changing curve of the curvature scalar, respectively. The blue solid curve is the coexistence curve. The blue dashed curve in the figure corresponds to the critical point temperature.}\label{R-THEE}
\end{figure}

In order to show the relation between the thermodynamic properties of the RN-AdS black hole and the holography entanglement entropy, we plot the characteristic curves with the Hawking temperature and entanglement entropy of CFT in Fig.\ref{R-THEE}.
The blue dashed line denotes the critical value of $t$, beyound which the corresponding black hole is undergoing a first-order phase transition. When the temperature is at the critical value, the black hole will show a second-order phase transition. What's more, the black hole is in the superficial phase $1/t$ exceed the blue dashed line. The blue solid curve is the coexistence curve, which separates into saturated small and large black hole phases by the critical point. The entanglement entropy $\delta S$ with a small or large value indicates that the black hole is thermodynamic stable. The orange curve is the divergent curve of the scalar curvature, which divided the unstable phase from the metastable phase. The red solid curve is the sign-changing curve, and the top area painted in red represents $\widetilde{\mathcal{R}}>0$. Furthermore, the correspondence of these characteristic curves between the black hole entropy and HEE announces that the thermodynamic information of the former can be read from the dual field theory.

\begin{figure}[!h]
	\centering
	\subfigure[]{
		\includegraphics[width=4.75cm]{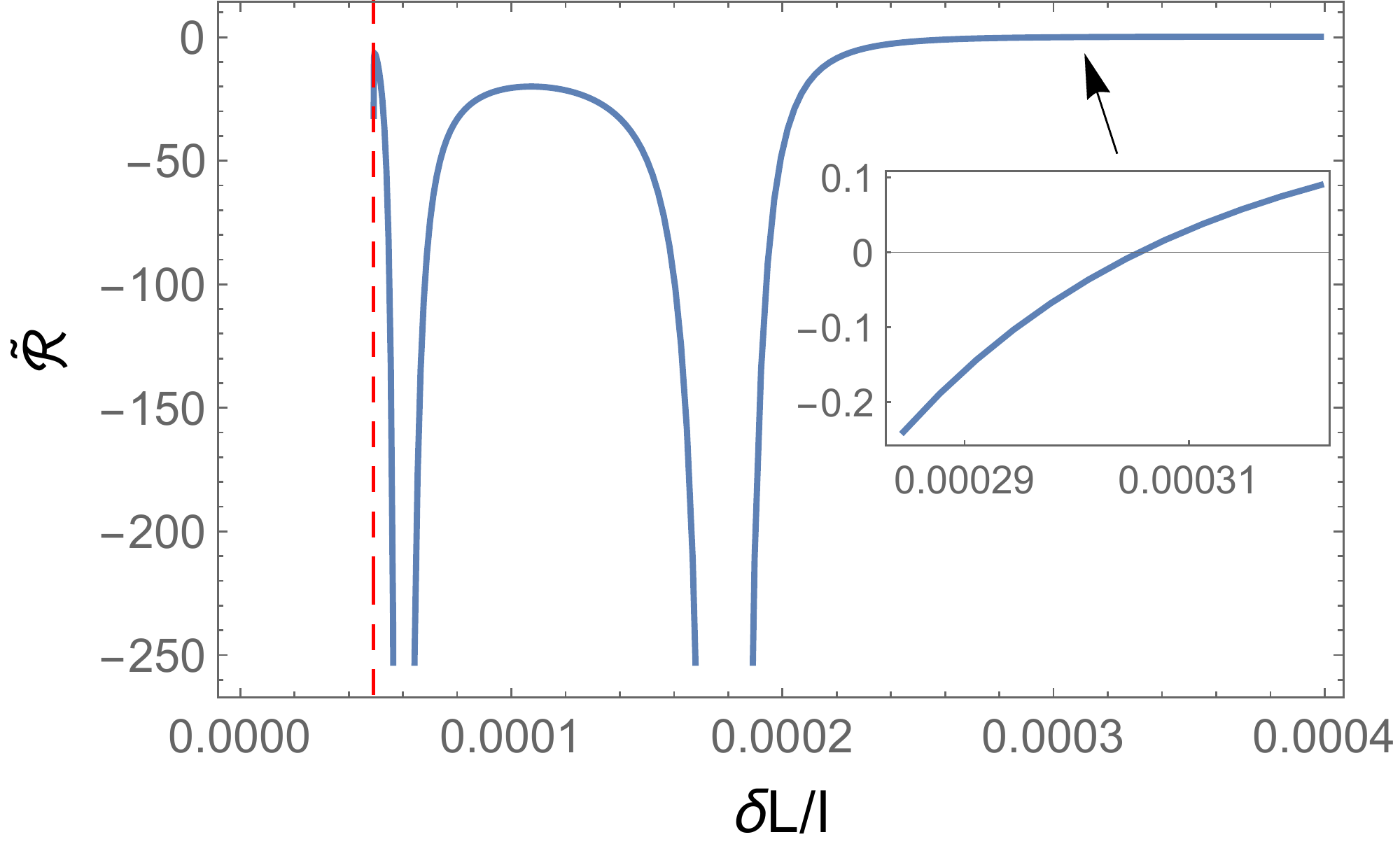}
	}
	\quad
	\subfigure[]{
		\includegraphics[width=4.75cm]{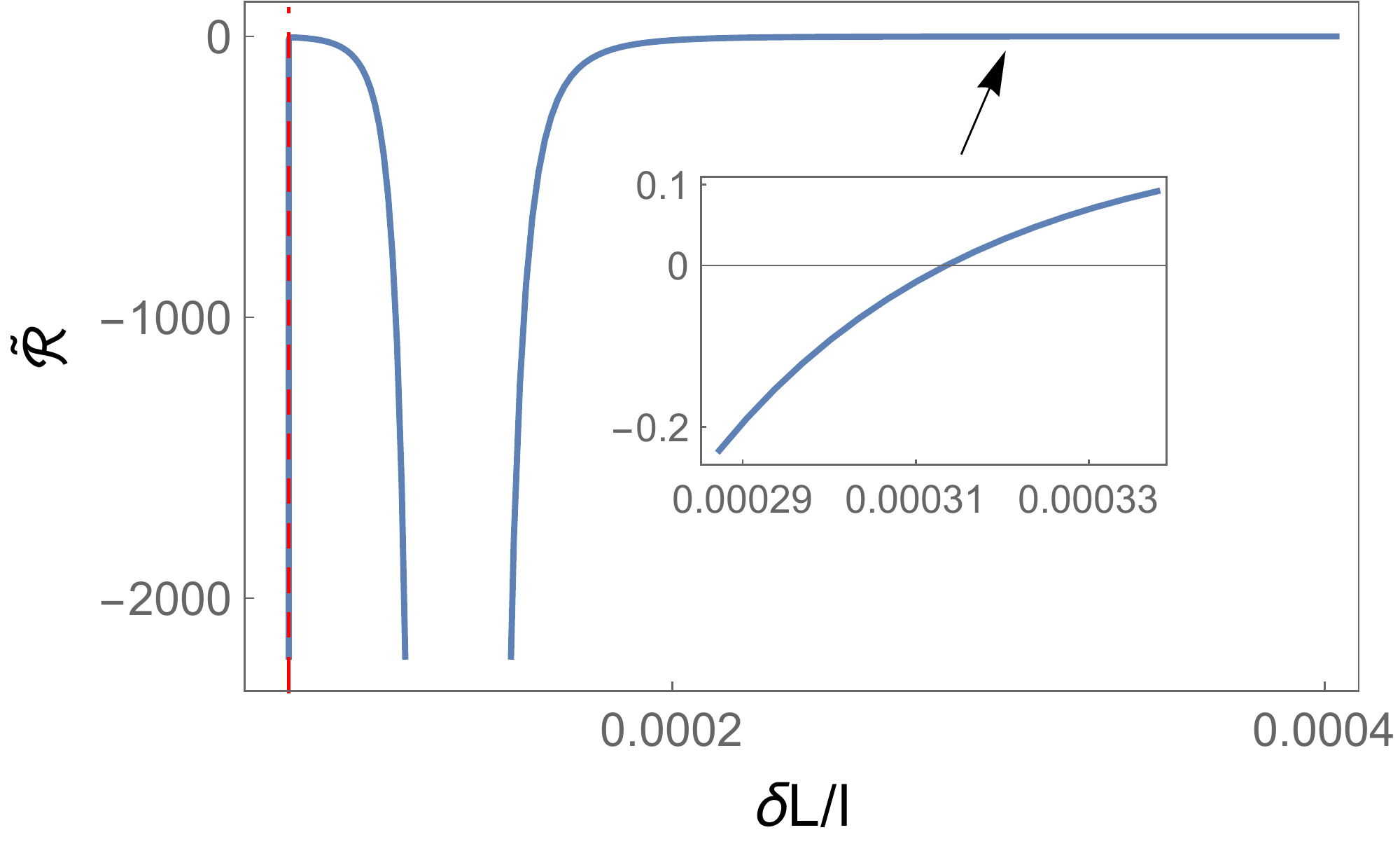}
	}
	\subfigure[]{
		\includegraphics[width=4.4cm]{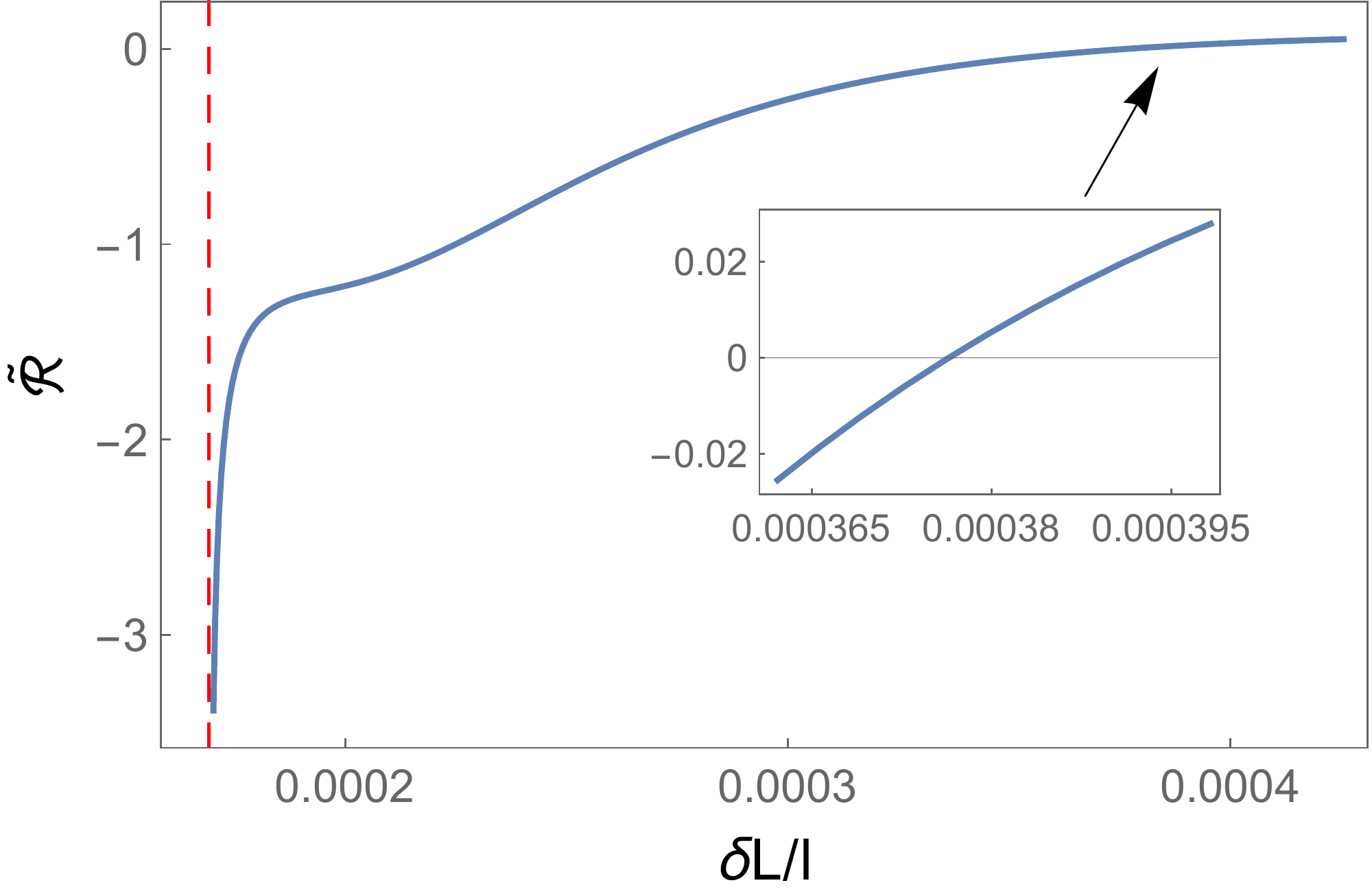}
	}
	\caption{The dimensionless curvature scalar for RN-AdS black hole in terms of $\delta L/l$. Here we set $q=0.6 q_c$, $q_c$ and $2q_c$ from left to right. The entropy on the red line correspond to the vanishing value of the reduced temperature $t$.}\label{R-dL}
\end{figure}

\begin{figure}[!h]
	\centering
	\includegraphics[width=7cm]{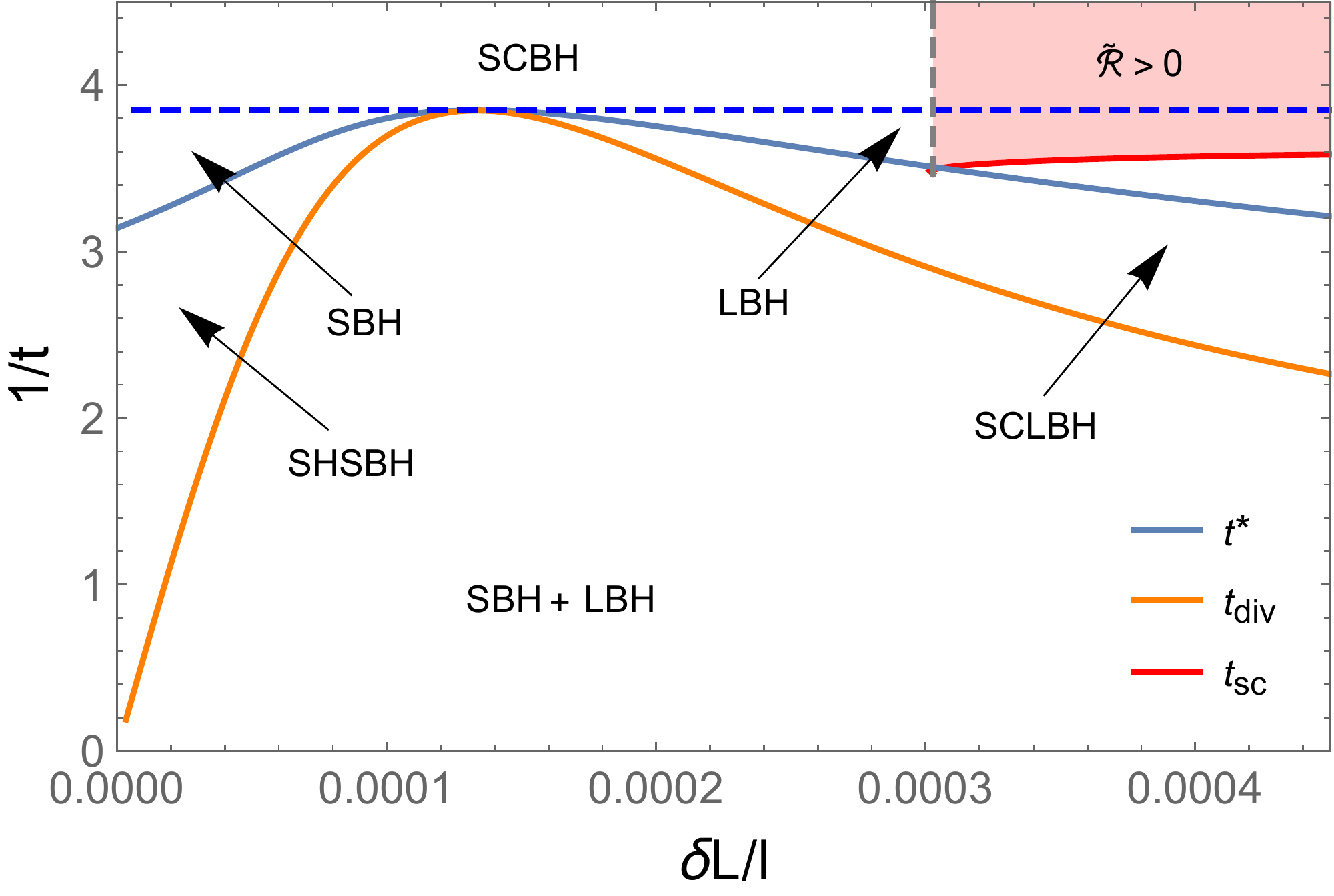}
	\caption{The characteristic curve of curvature scalar in coordinate space $\{ \delta L/l, 1/t\}$. The orange and red are solid lines are the divergence curve and the variable sign curve of the curvature scalar, respectively. The blue solid line is the coexistence curve. The blue dashed curve in the figure corresponds to the critical point temperature.}\label{R-L}
\end{figure}

Now, let's turn our attention to the two-point correlation function in conformal field theory. The AdS/CFT correspondence implied that the equal time two-point correlation function with large conformal dimension $\Delta$ of the scalar operator $\mathcal{O}$ of dual field theory is holographically approximated as  \cite{Balasubramanian:1999zv}
\begin{equation}
	< \mathcal{O} (t_0,x_i) \mathcal{O} (t_0,x_j)> \approx e^{-\Delta L} , \nonumber
\end{equation}
where $L$ is the length between the points $(t_0,x_i)$ and $(t_0,x_j)$ on the AdS boundary measured by the metric of the bulk geodesic. Due to the spacetime symmetry, we can let $x_i=\theta$, and the boundary is marked as $\theta_0$. So the proper length can be parameterized as 
\begin{equation}
	L=\int^{\theta_0}_\theta \mathcal{L}(r(\theta),r)\mathrm{d} \theta, \quad \mathcal{L}=\sqrt{\frac{(\dot{t})^2}{f(r)}+r^2},    \label{ActionT}
\end{equation}
here the dot denote $\dot{t}=\mathrm{d} r/\mathrm{d} \theta$. The equation of motion is obtained from the Euler-Lagrange equation with Lagrangian $\mathcal{L}$ with respect to $\theta$ . Applying the boundary condition Eq.(\ref{boundary}), $r(\theta)$ can be deduced by solving Eq.(\ref{ActionT}). We apply the numerical methods to calculate the geodesic length, which is difficult to obtain in analytical form. Due to the divergence of the geodesic length at the boundary $\theta_0$, it should be regularized by subtraction of the geodesic length in pure AdS with the same boundary condition, and will be denoted as $\delta L$. For that purpose, we choose $\theta_0=0.1$ and the UV-cutoff in the dual field theory as $r(0.099)$. The relationship between $\widetilde{\mathcal{R}}$ and $\delta L/l$ is shown in Fig.\ref{R-dL}.
The behavior of $\widetilde{\mathcal{R}}$ in Fig.\ref{R-dL} is a point-to-point correspondence with that in Fig.\ref{R-SQ} and Fig.\ref{R-HEE}. The result indicates that the underlying microstructure can also be probed by the two-point correlation function. To confirm this conclusion, we plot the characteristic curves in $\{ \delta L/l, 1/t \}$ in Fig.\ref{R-L}. 
As shown in the figure, the phase structure and the underlying microstructure are revealed with the quantities $\delta L$ and reduced temperature $t$ in the conformal field theory.

\section{Conclusion}\label{IIIII}

In this paper, we investigate the relationship between the thermodynamic geometry of the RN-AdS black hole and the non-local observables in a given CFT. In place of introducing the critical point associated with the black hole charge to non-dimensionalize the thermodynamic parameters, we use the cosmological constant to rescale the parameters, and this yields a universal black hole specific equation of state. Applying the black hole's internal energy, determined by the spacetime structure $U=M-\Phi Q$, we get the Ruppeiner curvature scalar Eq.(\ref{R}) and show the thermodynamic behavior of the black hole, which implies that $\widetilde{\mathcal{R} }$ can be exploited to display the phase transition of the black hole.

Through the connection of the black hole event horizon radius to non-local observers, the point-to-point correspondence of Ruppeiner geometry on black holes and dual fields is established. Below the critical charge, there are two divergent points divided $\widetilde{\mathcal{R} }$ into three-segment, implying that the black hole undergoes a first-order phase transition corresponding to the blue solid curve in Fig.\ref{T-S}. In this phase, the small stable black hole will directly jump to the large one as the temperature increase to coexistence temperature $t^*$. While as reduced temperature $t$ increases to the critical value, these two divergent points in Fig.\ref{R-HEE}(a) and Fig.\ref{R-L}(a) merge into one, and the phase transition becomes second-order. With the increasing of charge, the divergent point disappears, and the black hole was in the supercritical phase. In this process, the characteristic curve of $\widetilde{\mathcal{R} }$ show us the interaction in the black hole more clearly, which tells us that only the large cold black hole with large charge gets a positive value of curvature scalar, corresponding to a repulsive interaction domain inside the system. Our results implied that the observation of these non-local observables in dual field theory can reveals us the thermodynamic information of the AdS black hole.

In view of the invisibility of the black hole microstructure, if the holographic form of the thermodynamic geometry is established by the AdS/CFT dual theory, it may disclose the microscopic mechanism of the black hole thermodynamic behavior for us, and we will focus on this issue in our future work.

\section*{Acknowledgment}
This work is supported by the financial supports from the National Natural Science Foundation of China (Grant Nos.12275216, 12047502, 12105222), the China Postdoctoral Science Foundation (Grant Nos.2017M623219, 2020M673460). 
\end{spacing}

\providecommand{\href}[2]{#2}\begingroup
\footnotesize\itemsep=0pt
\providecommand{\eprint}[2][]{\href{http://arxiv.org/abs/#2}{arXiv:#2}}
\providecommand{\doi}[2]{\href{http://dx.doi.org/#2}{#1}}

\endgroup


\begin{thebibliography}{99}

\bibitem{Bekenstein:1973ur}
J.~D. Bekenstein, 
\emph{Black holes and entropy},
\href{https://doi.org/10.1103/PhysRevD.7.2333}{\emph{Phys. Rev. } {\bfseries D7} (1973) 2333--2346}.

\bibitem{Hawking:1974sw}
S.~W. Hawking, 
\emph{Particle Creation by Black Holes},
\href{https://doi.org/10.1007/BF02345020, 10.1007/BF01608497}{\emph{Commun. Math. Phys.} {\bfseries 43} (1975) 199--220}.

\bibitem{Hawking:1982dh}
S.~W.~Hawking and D.~N.~Page, 
\emph{Thermodynamics of Black Holes in anti-De Sitter Space},
\href{https://doi.org/10.1007/BF01208266}{\emph{Commun. Math. Phys. } {\bfseries 87} (1983) 577},

\bibitem{Witten:1998zw}
E.~Witten,
\emph{Anti-de Sitter space, thermal phase transition, and confinementin gauge theories},
\href{https://doi.org/10.4310/ATMP.1998.v2.n3.a3}{\emph{Adv. Theor. Math. Phys. } {\bfseries 2} (1998) 505--532},
[\href{https://arxiv.org/abs/hep-th/9803131}{{\ttfamily arXiv:hep-th/9803131}}].

\bibitem{Chamblin:1999tk}
A.~Chamblin, R.~Emparan, C.~V.~Johnson and R.~C.~Myers,
\emph{Charged AdS black holes and catastrophic holography},
\href{https://doi.org/10.1103/PhysRevD.60.064018}{\emph{Phys. Rev. } {\bfseries D60} (1999) 064018}.
[\href{https://arxiv.org/abs/hep-th/9902170}{{\ttfamily arXiv:hep-th/9902170}}].

\bibitem{Kastor:2009wy}
D.~Kastor, S.~Ray and J.~Traschen, 
\emph{Enthalpy and the Mechanics of AdS Black Holes}, 
\href{https://doi.org/10.1088/0264-9381/26/19/195011}{\emph{Class. Quant. Grav. } {\bfseries 26} (2009) 195011},
[\href{https://arxiv.org/abs/0904.2765}{{\ttfamily arXiv:0904.2765}}].

\bibitem{Dolan:2011xt}
B.~P. Dolan, 
\emph{Pressure and volume in the first law of black hole thermodynamics},
\href{https://doi.org/10.1088/0264-9381/28/23/235017}{\emph{Class. Quant. Grav. } {\bfseries 28} (2011) 235017},
[\href{https://arxiv.org/abs/1106.6260}{{\ttfamily arXiv:1106.6260}}].

\bibitem{Cvetic:2010jb}
M.~Cvetic, G.~W. Gibbons, D.~Kubiznak and C.~N. Pope, 
\emph{Black Hole Enthalpy and an Entropy Inequality for the Thermodynamic Volume},
\href{https://doi.org/10.1103/PhysRevD.84.024037}{\emph{Phys. Rev. }{\bfseries D84} (2011) 024037},
[\href{https://arxiv.org/abs/1012.2888}{{\ttfamily arXiv:1012.2888}}].

\bibitem{Kubiznak:2012wp}
D.~Kubiznak and R.~B.~Mann, 
\emph{P-V criticality of charged AdS black holes}, 
\href{https://doi.org/10.1007/JHEP07(2012)033}{\emph{JHEP } {\bfseries 07} (2012) 033},
[\href{https://arxiv.org/abs/1205.0559}{{\ttfamily arXiv:1205.0559}}].

\bibitem{Kubiznak:2016qmn}
D.~Kubiznak, R.~B. Mann and M.~Teo, 
\emph{Black hole chemistry: thermodynamics with Lambda}, 
\href{https://doi.org/10.1088/1361-6382/aa5c69}{\emph{Class. Quant. Grav. } {\bfseries 34} (2017) 063001},
[\href{https://arxiv.org/abs/1608.06147}{{\ttfamily arXiv:1608.06147}}].

\bibitem{Toledo:2019amt}
J.~M. Toledo, and V.~B. Bezerra,
\emph{Some remarks on the thermodynamics of charged AdS black holes with cloud of strings and quintessence},
\href{https://doi.org/10.1140/epjc/s10052-019-6616-8}{\emph{Eur. Phys. J. }{\bfseries C79} (2019) 110}.

\bibitem{Hendi:2012um}
S.~H. Hendi and M.~H. Vahidinia, 
\emph{Extended phase space thermodynamics and P-V criticality of black holes with a nonlinear source},
\href{https://doi.org/10.1103/PhysRevD.88.084045}{\emph{Phys. Rev. }{\bfseries D88}  (2013) 084045},
[\href{https://arxiv.org/abs/1212.6128}{{\ttfamily arXiv:1212.6128}}].

\bibitem{Wei:2012ui}
S.-W. Wei and Y.-X. Liu, 
\emph{Critical phenomena and thermodynamic geometry of charged Gauss-Bonnet AdS black holes},
\href{https://doi.org/10.1103/PhysRevD.87.044014}{\emph{Phys. Rev. }{\bfseries D87}  (2013) 044014},
[\href{https://arxiv.org/abs/1209.1707}{{\ttfamily arXiv:1209.1707}}].

\bibitem{Cai:2013qga}
R.-G. Cai, L.-M. Cao, L.~Li and R.-Q. Yang, 
\emph{P-V criticality in the extended phase space of Gauss-Bonnet black holes in AdS space},
\href{https://doi.org/10.1007/JHEP09(2013)005}{\emph{JHEP } {\bfseries 09}  (2013) 005}, 
[\href{https://arxiv.org/abs/1306.6233}{{\ttfamily arXiv:1306.6233}}].

\bibitem{Zhao:2013oza}
R.~Zhao, H.-H. Zhao, M.-S. Ma and L.-C. Zhang, 
\emph{On the critical phenomena and thermodynamics of charged topological dilaton AdS black holes},
\href{https://doi.org/10.1140/epjc/s10052-013-2645-x}{\emph{Eur. Phys. J. }{\bfseries C73} (2013) 2645},
[\href{https://arxiv.org/abs/1305.3725}{{\ttfamily arXiv:1305.3725}}].

\bibitem{Altamirano:2013ane}
N.~Altamirano, D.~Kubiznak and R.~B. Mann, 
\emph{Reentrant phase transitions in rotating anti-de Sitter black holes},
\href{https://doi.org/10.1103/PhysRevD.88.101502}{\emph{Phys. Rev. }{\bfseries D88} (2013) 101502},
[\href{https://arxiv.org/abs/1306.5756}{{\ttfamily arXiv:1306.5756}}].

\bibitem{Spallucci:2013osa}
E.~Spallucci and A.~Smailagic, 
\emph{Maxwell's equal area law for charged Anti-de Sitter black holes},
\href{https://doi.org/10.1016/j.physletb.2013.05.038}{\emph{Phys. Lett. }{\bfseries B723} (2013) 436--441},
[\href{https://arxiv.org/abs/1305.3379}{{\ttfamily arXiv:1305.3379}}].

\bibitem{Xu:2014tja}
H.~Xu, W.~Xu and L.~Zhao, 
\emph{Extended phase space thermodynamics for third order Lovelock black holes in diverse dimensions},
\href{https://doi.org/10.1140/epjc/s10052-014-3074-1}{\emph{Eur. Phys. J. }{\bfseries C74} (2014) 3074},
[\href{https://arxiv.org/abs/1405.4143}{{\ttfamily arXiv:1405.4143}}].

\bibitem{Maldacena:1997re}
J.~M.~Maldacena,
\emph{The Large N limit of superconformal field theories and supergravity},
\href{https://doi.org/10.1023/A:1026654312961}{\emph{Adv. Theor. Math. Phys. }{\bfseries 2} (1998) 231-252},
[\href{https://arxiv.org/abs/hep-th/9711200}{{\ttfamily arXiv:hep-th/9711200}}].

\bibitem{Gubser:1998bc}
S.~S.~Gubser, I.~R.~Klebanov and A.~M.~Polyakov,
\emph{Gauge theory correlators from noncritical string theory},
\href{https://doi.org/10.1016/S0370-2693(98)00377-3}{\emph{Phys. Lett. }{\bfseries B428} (1998) 105-114},
[\href{https://arxiv.org/abs/hep-th/9802109}{{\ttfamily arXiv:hep-th/9802109}}].

\bibitem{Dolan:2014cja}
B.~P.~Dolan,
\emph{Bose condensation and branes},
\href{https://doi.org/10.1007/JHEP10(2014)179}{\emph{JHEP }{\bfseries 10} (2014) 179},
[\href{https://arxiv.org/abs/1406.7267}{{\ttfamily arXiv:1406.7267}}].

\bibitem{Kastor:2014dra}
D.~Kastor, S.~Ray and J.~Traschen,
\emph{Chemical Potential in the First Law for Holographic Entanglement Entropy},
\href{https://doi.org/10.1007/JHEP11(2014)120}{\emph{JHEP }{\bfseries 11} (2014) 120},
[\href{https://arxiv.org/abs/1409.3521}{{\ttfamily arXiv:1409.3521}}].

\bibitem{Karch:2015rpa}
A.~Karch and B.~Robinson,
\emph{Holographic Black Hole Chemistry},
\href{https://doi.org/10.1007/JHEP12(2015)073}{\emph{JHEP }{\bfseries 12} (2015) 073},
[\href{https://arxiv.org/abs/1510.02472}{{\ttfamily arXiv:1510.02472}}].

\bibitem{Henningson:1998gx}
M.~Henningson and K.~Skenderis,
\emph{The Holographic Weyl anomaly},
\href{https://doi.org/10.1088/1126-6708/1998/07/023}{\emph{JHEP }{\bfseries 07}, (1998) 023},
[\href{https://arxiv.org/abs/hep-th/9806087}{{\ttfamily arXiv:hep-th/9806087}}].

\bibitem{Freedman:1999gp}
D.~Z.~Freedman, S.~S.~Gubser, K.~Pilch and N.~P.~Warner,
\emph{Renormalization group flows from holography supersymmetry and a c theorem},
\href{https://doi.org/10.4310/ATMP.1999.v3.n2.a7}{\emph{Adv. Theor. Math. Phys. }{\bfseries 3}, (1999) 363-417},
[\href{https://arxiv.org/abs/hep-th/9904017}{{\ttfamily arXiv:hep-th/9904017}}].

\bibitem{Myers:2010xs}
R.~C.~Myers and A.~Sinha,
\emph{Seeing a c-theorem with holography},
\href{https://doi.org/10.1103/PhysRevD.82.046006}{\emph{Phys. Rev. }{\bfseries D82}, (2010) 046006},
[\href{https://arxiv.org/abs/1006.1263}{{\ttfamily arXiv:1006.1263}}].

\bibitem{Cong:2021fnf}
W.~Cong, D.~Kubiznak and R.~B.~Mann,
\emph{Thermodynamics of AdS Black Holes: Critical Behavior of the Central Charge},
\href{https://doi.org/10.1103/PhysRevLett.127.091301}{\emph{Phys. Rev. Lett. }{\bfseries 127}, (2021) 091391},
[\href{https://arxiv.org/abs/2105.02223}{{\ttfamily arXiv:2105.02223}}].

\bibitem{Visser:2021eqk}
M.~R.~Visser, 
\emph{Holographic Thermodynamics Requires a Chemical Potential for Color},
\href{https://doi.org/10.1103/PhysRevD.105.106014}{\emph{Phys. Rev. }{\bfseries D105}, (2022) 106014},
[\href{https://arxiv.org/abs/2101.04145}{{\ttfamily arXiv:2101.04145}}].

\bibitem{Rafiee:2021hyj}
M.~Rafiee, S.~A.~H.~Mansoori, S.~W.~Wei and R.~B.~Mann,
\emph{Universal criticality of thermodynamic geometry for boundary conformal field theories in gauge/gravity duality},
\href{https://doi.org/10.1103/PhysRevD.105.024058}{\emph{Phys. Rev. }{\bfseries D105}, (2022) 024058},
[\href{https://arxiv.org/abs/2107.08883}{{\ttfamily arXiv:2107.08883}}].

\bibitem{Cong:2021jgb}
W.~Cong, D.~Kubiznak, R.~Mann and M.~Visser,
\emph{Holographic CFT Phase Transitions and Criticality for Charged AdS Black Holes},
[\href{https://arxiv.org/abs/2112.14848}{{\ttfamily arXiv:2112.14848}}].

\bibitem{Zeyuan:2021uol}
G.~Zeyuan and L.~Zhao,
\emph{Restricted phase space thermodynamics for AdS black holes via holography},
\href{https://doi.org/10.1088/1361-6382/ac566c}{\emph{Class. Quant. Grav. }{\bfseries 39}, (2022) 075019},
[\href{https://arxiv.org/abs/2112.02386}{{\ttfamily arXiv:2112.02386}}].

\bibitem{Gao:2021xtt}
Z.~Y.~Gao, X.~Q.~Kong and L.~Zhao,
\emph{Thermodynamics of Kerr-AdS black holes in the restricted phase space},
\href{https://doi.org/10.1140/epjc/s10052-022-10080-y}{\emph{Eur. Phys. J. }{\bfseries C82}, (2022) 112},
[\href{https://arxiv.org/abs/2112.08672}{{\ttfamily arXiv:2112.08672}}].

\bibitem{Wang:2021cmz}
T.~Wang and L.~Zhao,
\emph{Black hole thermodynamics is extensive with variable Newton constant},
\href{https://doi.org/10.1016/j.physletb.2022.136935}{\emph{Phys. Lett. }{\bfseries B827},  (2022) 136935},
[\href{https://arxiv.org/abs/2112.11236}{{\ttfamily arXiv:2112.11236}}].

\bibitem{Zhao:2022dgc}
L.~Zhao,
\emph{Thermodynamics for higher dimensional rotating black holes with variable Newton constant},
\href{https://doi.org/10.1088/1674-1137/ac4f4c}{\emph{Chin. Phys. }{\bfseries C46},  (2022) 055105},
[\href{https://arxiv.org/abs/2201.00521}{{\ttfamily arXiv:2201.00521}}].

\bibitem{Johnson:2013dka}
C.~V.~Johnson,
\emph{Large N Phase Transitions, Finite Volume, and Entanglement Entropy},
\href{https://doi.org/10.1007/JHEP03(2014)047}{\emph{JHEP }{\bfseries 03} (2014) 047},
[\href{https://arxiv.org/abs/1306.4955}{{\ttfamily arXiv:1306.4955}}].

\bibitem{Ryu:2006bv}
S.~Ryu and T.~Takayanagi,
\emph{Holographic derivation of entanglement entropy from AdS/CFT},
\href{https://doi.org/10.1103/PhysRevLett.96.181602}{\emph{Phys. Rev. Lett. }{\bfseries 96} (2006) 181602},
[\href{https://arxiv.org/abs/hep-th/0603001}{{\ttfamily arXiv:hep-th/0603001}}].

\bibitem{Ryu:2006ef}
S.~Ryu and T.~Takayanagi,
\emph{Aspects of Holographic Entanglement Entropy},
\href{https://doi.org/10.1088/1126-6708/2006/08/045}{\emph{JHEP }{\bfseries 08} (2006) 045},
[\href{https://arxiv.org/abs/hep-th/0605073}{{\ttfamily arXiv:hep-th/0605073}}].

\bibitem{Caceres:2015vsa}
E.~Caceres, P.~H.~Nguyen and J.~F.~Pedraza,
\emph{Holographic entanglement entropy and the extended phase structure of STU black holes},
\href{https://doi.org/10.1007/JHEP09(2015)184}{\emph{JHEP }{\bfseries 09} (2015) 184},
[\href{https://arxiv.org/abs/1507.06069}{{\ttfamily arXiv:1507.06069}}].

\bibitem{Nguyen:2015wfa}
P.~H.~Nguyen,
\emph{An equal area law for holographic entanglement entropy of the AdS-RN black hole},
\href{https://doi.org/10.1007/JHEP12(2015)139}{\emph{JHEP }{\bfseries 12} (2015) 139},
[\href{https://arxiv.org/abs/1508.01955}{{\ttfamily arXiv:1508.01955}}].

\bibitem{Zeng:2015wtt}
X.~X.~Zeng and L.~F.~Li,
\emph{Van der Waals phase transition in the framework of holography},
\href{https://doi.org/10.1016/j.physletb.2016.11.017}{\emph{Phys. Lett. }{\bfseries B764} (2017) 100-108},
[\href{https://arxiv.org/abs/1512.08855}{{\ttfamily arXiv:1512.08855}}].

\bibitem{Zeng:2016fsb}
X.~X.~Zeng and L.~F.~Li,
\emph{Holographic Phase Transition Probed by Nonlocal Observables},
\href{https://doi.org/10.1155/2016/6153435}{\emph{Adv. High Energy Phys. }{\bfseries 2016} (2016) 6153435},
[\href{https://arxiv.org/abs/1609.06535}{{\ttfamily arXiv:1609.06535}}].

\bibitem{PhysRevA.24.488}
G.~Ruppeiner, 
\emph{Application of Riemannian geometry to the thermodynamics of a simple fluctuating magnetic system},
\href{https://doi.org/10.1103/PhysRevA.24.488}{\emph{Phys. Rev. } {\bfseries A24} (1981) 488--492}.

\bibitem{Ruppeiner:1983zz}
G.~Ruppeiner, 
\emph{Thermodynamic Critical Fluctuation Theory?}
\href{https://doi.org/10.1103/PhysRevLett.50.287}{\emph{Phys. Rev. Lett. }{\bfseries 50} (1983) 287--290}.

\bibitem{Ruppeiner:1995zz}
G.~Ruppeiner, 
\emph{Riemannian geometry in thermodynamic fluctuation theory},
\href{https://doi.org/10.1103/RevModPhys.67.605}{\emph{Rev. Mod. Phys. }{\bfseries 67} (1995) 605--659}.

\bibitem{Ruppeiner:2008kd}
G.~Ruppeiner,
\emph{Thermodynamic curvature and phase transitions in Kerr-Newman black holes},
\href{https://doi.org/10.1103/PhysRevD.78.024016}{\emph{Phys. Rev. }{\bfseries D78} (2008) 024016},
[\href{https://arxiv.org/abs/0802.1326}{{\ttfamily arXiv:0802.1326}}].


\bibitem{Ruppeiner:2010}
G.~Ruppeiner,
\emph{Thermodynamic curvature measures interactions},
\href{http://dx.doi.org/10.1119/1.3459936}{\emph{Am. J. Phys. }{\bfseries 78} (2010) 1107},
[\href{https://arxiv.org/abs/1007.2160}{{\ttfamily arXiv:1007.2160}}].


\bibitem{Sahay:2010wi}
A.~Sahay, T.~Sarkar and G.~Sengupta,
\emph{Thermodynamic Geometry and Phase Transitions in Kerr-Newman-AdS Black Holes},
\href{https://doi.org/10.1007/JHEP04(2010)118}{\emph{JHEP }{\bfseries 04} (2010) 118},
[\href{https://arxiv.org/abs/1002.2538}{{\ttfamily arXiv:1002.2538}}].

\bibitem{Wei:2015iwa}
S.~W.~Wei and Y.~X.~Liu,
\emph{Insight into the Microscopic Structure of an AdS Black Hole from a Thermodynamical Phase Transition},
\href{https://doi.org/10.1103/PhysRevLett.115.111302}{\emph{Phys. Rev. Lett. }{\bfseries 116} (2016) 169903},
[\href{https://arxiv.org/abs/1502.00386}{{\ttfamily arXiv:1502.00386}}].

\bibitem{Zhang:2015ova}
J.~L.~Zhang, R.~G.~Cai and H.~Yu,
\emph{Phase transition and thermodynamical geometry of Reissner-Nordstr\"om-AdS black holes in extended phase space},
\href{https://doi.org/10.1103/PhysRevD.91.044028}{\emph{Phys. Rev. }{\bfseries D91} (2015) 044028},
[\href{https://arxiv.org/abs/1502.01428}{{\ttfamily arXiv:1502.01428}}].


\bibitem{Wei:2017icx}
S.-W. Wei, B.~Liang and Y.-X. Liu, 
\emph{Critical phenomena and chemical potential of a charged AdS black hole},
\href{https://doi.org/10.1103/PhysRevD.96.124018}{\emph{Phys. Rev. }{\bfseries D96} (2017) 124018},
[\href{https://arxiv.org/abs/1705.08596}{{\ttfamily arXiv:1705.08596}}].

\bibitem{Chaturvedi:2017vgq}
P.~Chaturvedi, S.~Mondal and G.~Sengupta, 
\emph{Thermodynamic Geometry of Black Holes in the Canonical Ensemble},
\href{https://doi.org/10.1103/PhysRevD.98.086016}{\emph{Phys. Rev. }{\bfseries D98} (2018) 086016},
[\href{https://arxiv.org/abs/1705.05002}{{\ttfamily arXiv:1705.05002}}].

\bibitem{Wei:2019uqg}
S.~W.~Wei, Y.~X.~Liu and R.~B.~Mann,
\emph{Repulsive Interactions and Universal Properties of Charged Anti\textendash{}de Sitter Black Hole Microstructures},
\href{https://doi.org/10.1103/PhysRevLett.123.071103}{\emph{Phys. Rev. Lett. }{\bfseries 123} (2019) 071103},
[\href{https://arxiv.org/abs/1906.10840}{{\ttfamily arXiv:1906.10840}}].

\bibitem{Wei:2019yvs}
S.~W.~Wei, Y.~X.~Liu and R.~B.~Mann,
\emph{Ruppeiner Geometry, Phase Transitions, and the Microstructure of Charged AdS Black Holes},
\href{https://doi.org/10.1103/PhysRevD.100.124033}{\emph{Phys. Rev. D }{\bfseries 100} (2019) 124033},
[\href{https://arxiv.org/abs/1909.03887}{{\ttfamily arXiv:1909.03887}}].

\bibitem{Xu:2020ftx}
Z.~M.~Xu, B.~Wu and W.~L.~Yang,
\emph{Diagnosis inspired by the thermodynamic geometry for different thermodynamic schemes of the charged BTZ black hole},
\href{https://doi.org/10.1140/epjc/s10052-020-08563-x}{\emph{Eur. Phys. J. }{\bfseries C80} (2020) 997 (2020)},
[\href{https://arxiv.org/abs/2002.00117}{{\ttfamily arXiv:2002.00117}}].

\bibitem{Wang:2021vbn}
C.~Wang, B.~Wu, Z.~M.~Xu and W.~L.~Yang,
\emph{Ruppeiner geometry of the RN-AdS black hole using shadow formalism},
\href{https://doi.org/10.1016/j.nuclphysb.2022.115698}{\emph{Nucl. Phys. }{\bfseries B976} (2022) 115698},
[\href{https://arxiv.org/abs/2107.12615}{{\ttfamily arXiv:2107.12615}}].

\bibitem{Wang:2021llu}
P.~Wang and F.~Y,~Yao,
\emph{Thermodynamic geometry of black holes enclosed by a cavity in extended phase space},
\href{https://doi.org/10.1016/j.nuclphysb.2022.115715}{\emph{Nucl. Phys. }{\bfseries B976} (2022) 115715},
[\href{https://arxiv.org/abs/2107.14640}{{\ttfamily arXiv:2107.14640}}].

\bibitem{Ruppeiner:2011gm}
G.~Ruppeiner, A.~Sahay, T.~Sarkar and G.~Sengupta, 
\emph{Thermodynamic Geometry, Phase Transitions, and the Widom Line},
\href{https://doi.org/10.1103/PhysRevE.86.052103}{\emph{Phys. Rev. }{\bfseries E86} (2012) 052103},
[\href{https://arxiv.org/abs/1106.2270}{{\ttfamily arXiv:1106.2270}}].

\bibitem{Shen:2005nu}
J.~Y.~Shen, R.~G.~Cai, B.~Wang and R.~K.~Su,
\emph{Thermodynamic geometry and critical behavior of black holes},
\href{https://doi.org/10.1142/S0217751X07034064}{\emph{Int. J. Mod. Phys. }{\bfseries A22} (2007) 11-27},
[\href{https://arxiv.org/abs/gr-qc/0512035}{{\ttfamily arXiv:gr-qc/0512035}}].

\bibitem{Balasubramanian:1999zv}
V.~Balasubramanian and S.~F.~Ross,
\emph{Holographic particle detection},
\href{https://doi.org/10.1103/PhysRevD.61.044007}{\emph{Phys. Rev. }{\bfseries D61} (2000) 044007},
[\href{https://arxiv.org/abs/hep-th/9906226}{{\ttfamily arXiv:hep-th/9906226}}].









\end{thebibliography}
\end{document}